\documentclass[useAMS,usenatbib,usegraphicx,letterpaper]{mn2e}

\newcommand{\degt}{deg$^2$}

\title[Discovery of a luminous $\boldmath z=5.9$ quasar in the UKIDSS LAS]
{The discovery of the first luminous $\boldmath z\sim6$ quasar in the UKIDSS  Large Area Survey} 
\author[B. P. Venemans, et al.]{B. P. Venemans$^{1}$ \thanks{E-mail:
venemans@ast.cam.ac.uk}, R. G. McMahon$^{1}$, S. J. Warren$^{2}$, 
E. A. Gonzalez-Solares$^{1}$, 
\newauthor
P. C. Hewett$^{1}$, D. J. Mortlock$^{2}$, S. Dye$^{3}$, R. G. Sharp$^{4}$
\\
$^{1}$ Institute of Astronomy, University of Cambridge, Madingley Road, Cambridge, CB3 0HA, United Kingdom\\
$^{2}$ Astrophysics Group, Imperial College London, Blackett Laboratory, Prince Consort Road, London, SW7 2AZ, United Kingdom\\
$^{3}$ Cardiff University, School of Physics \& Astronomy, Queens Buildings, The Parade, Cardiff, CF24 3AA, United Kingdom\\
$^{4}$ Anglo-Australian Observatory, P.O. Box 296, Epping, NSW 1710, Australia\\}

\begin{document}

\date{Accepted 2007 January 12. Received 2007 January 10; in original form 2006 December 6}

\pagerange{\pageref{firstpage}--\pageref{lastpage}} \pubyear{2006}

\maketitle

\label{firstpage}

\begin{abstract}
  We present the initial results from our search for high redshift, $z
  \ga 6$, quasars using near infrared data from the UKIRT Infrared
  Deep Sky Survey (UKIDSS) Large Area Survey (LAS). Our analysis of
  106 deg$^2$ of sky from Data Release 1 (DR1) has resulted in the
  discovery of ULAS J020332.38+001229.2, a luminous ($J_{\rm
    AB}=20.0$, $J_{\rm Vega}=19.1$, $M_{1450}=-26.2$) quasar at
  $z=5.86$. Following candidate selection from the combined IR and
  optical catalogue data and stacking of multiple epoch Sloan Digital
  Sky Survey (SDSS) data, we have obtained optical spectroscopy for
  the only two $z \ga 6$ quasar candidates. The VLT FORS2 spectrum of
  ULAS J020332.38+001229.2 shows broad Ly$\alpha$ +
  N\,{\sevensize{V}}~1240 emission at $\sim$8350\,\AA\ and an abrupt
  continuum break due to absorption by the Ly$\alpha$ forest. The
  quasar is not present in the SDSS DR5 catalogue and the continuum
  spectral index of $\alpha$=-1.4 ($F_{\nu}\propto\nu^{\alpha}$) is
  redder than a composite of SDSS quasars at similar redshifts
  ($\alpha$=-0.5). The discovery of one $z\sim6$ quasar in $\sim$100
  \degt\ in a complete sample within our selection criteria down to a
  median depth of $Y_{\rm AB}=20.4$ (7$\sigma$) is consistent with
  existing SDSS results. We describe our survey methodology,
  including the use of optical data from the SDSS and the highly
  effective procedures developed to isolate the very small surface
  density of high-probability quasar candidates.
\end{abstract}


\section{Introduction}

Quasars are powerful probes of the high-redshift Universe since, by
virtue of their high luminosities, they can be observed out to immense
distances with large look back times. The host galaxies of the highest
redshift quasars are probably still forming and most likely occur in
overdense regions in the matter distribution of the early
Universe. Studies of their space densities, host galaxies and local
environments provide potentially powerful constraints on theories of
the formation and growth of supermassive black holes, the growth of
large scale structure and the formation and evolution of the first
galaxies (e.g.\ Kauffmann \& Haehnelt 2000).

In addition to being of intrinsic interest, bright high-redshift
quasars are particularly valuable as probes of the Universe via
absorption studies of cosmologically distributed intervening material
and can be used to determine the baryonic content and physical
conditions (metallicity, temperature, ionization state) of the
intergalactic medium (IGM) and gas within intervening galaxies.

Finding quasars above redshift $z>6.0$ is of crucial importance since
it is in the epoch of reionization between the redshift where the
Str\"omgren spheres of the reionization sources overlap
($z_{\mathrm{OVL}}=6.1\pm0.15$, Gnedin \& Fan 2006), as shown by the
observations of the Gunn-Peterson optical depth at $z\sim6$, and the
lower bound (1$\sigma$) to the epoch of reionization from the 3-year
WMAP polarization results at $z=8.2$ (Page et al.\ 2006) which
corresponds roughly to the epoch when the Universe was half-ionized
(by volume).

Over the last few years there has been considerable progress in wide
area searches for high-redshift quasars. Storrie-Lombardi et al.\
(2001) reported the final results of a large survey for $z\sim4$
quasars over $\sim$8000\,deg$^2$ which resulted in $\sim$50 quasars in
the redshift range $4.0 < z < 4.8$. This survey used photographic data
and was based on the drop across Ly$\alpha$ caused by the Ly$\alpha$
forest, depressing the observed flux in the $B$-band relative to the
$R$-band. The advent of the CCD-based Sloan Digital Sky Survey (SDSS,
York et al.\ 2000) extended wide-area surveys to longer wavelengths
with the inclusion of the $i$- and $z$-bands, enabling searches for
quasars based on $i-z$ colour and redshifts of $z\sim6$ were reached
(Fan et al.\ 2000). There are now around 20 $z\sim6$ quasars known,
with the highest redshift quasar at $z=6.42$ (Fan et al.\ 2003, 2006).

However, finding quasars beyond $z=6.4$ using the SDSS is practically
almost impossible because the sources become too faint in the reddest
passband (the $z$-band) due to absorption by the intervening
Ly$\alpha$ forest. To reach higher redshift requires wide-field
near-infrared imaging to substantially fainter fluxes then the 2
Micron All Sky Survey (2MASS, Skrutskie et al.\ 2006). This is the
parameter space explored by the UK Infrared Telescope (UKIRT) Infrared
Deep Sky Survey (UKIDSS) Large Area Survey (LAS). One of the main
science drivers of the UKIDSS LAS is the discovery of quasars with
$z>6$.

In this Letter, we report the first high-redshift quasar discovered
with UKIDSS, ULAS J020332.38+001229.2 at $z=5.86$. All magnitudes are
given in the AB system using the Vega to AB conversions provided in
Hewett et al.\ (2006) and a $\Lambda$-dominated cosmology with H$_0 =
70$ km\,s$^{-1}$\,Mpc$^{-1}$, $\Omega_{M} = 0.3$, and
$\Omega_{\Lambda} = 0.7$ is adopted.

\section{Survey imaging data}

The near infrared data are from the UKIDSS LAS (Lawrence et al.\ 2006)
which is being carried out using the UKIRT Wide Field Camera (WFCAM,
Casali et al.\ 2006) on the 3.8-m UKIRT. WFCAM has a sparse packed
mosaic of four 2048$\times$2048 pixel Rockwell Hawaii-II arrays with a
pixel scale of 0\farcs4 pixel$^{-1}$. The instantaneous field of view
is 0.21 \degt. The arrays are spaced by 0.94 times the detector
width. A mosaic of four pointings gives a tile with a continuous
covered area of 0.77 \degt. See Dye et al.\ (2006) for technical
details of UKIDSS.

The LAS consists of nominal 40-s exposures in the $Y$-, $J$-, $H$- and
$K$-bands. The $Y$-band (0.97--1.07\,$\mu$m) is specially designed to
fall between the $z$-band and the $J$-band, occupying the clean
wavelength range between the atmospheric absorption bands at
0.95\,$\mu$m and 1.14\,$\mu$m. The $Y$-band is needed to distinguish
high-redshift quasars from the more numerous cool stars, such as L-
and T-dwarfs (Hewett et al.\ 2006, see also Fig.\ \ref{iyj}).

In this Letter, we use data from the Data Release 1 (DR1, Warren et
al.\ 2006). The UKIDSS observations were obtained during the period
2005 August to 2006 January. The data was released to the ESO
community in 2006 July (see Warren et al.\ 2006 for further
details). The median 5$\sigma$ point source depths in AB magnitudes
are $Y_{\rm AB}$=20.8, $J_{\rm AB}$=20.5, $H_{\rm AB}$=20.2 and
$K_{\rm AB}$=20.1 (Warren et al.\ 2006). We have restricted our
analysis to the UKIDSS LAS area that is imaged at multiple epochs by
the SDSS Southern Survey ($\rm -25^\circ < RA < 60^\circ; -1.25^\circ
< Dec < 1.27^\circ$, called Stripe 82 in the SDSS footprint, York et
al.\ 2000). This region of sky is being imaged by the SDSS
collaboration in 5 passbands ($u$, $g$, $r$, $i$, $z$) repeatedly
during three months (September, October and November) in each of three
years (2005-2007) as part of the SDSS-II project. In addition Stripe
82 was imaged 10-20 times during the SDSS-I project (see
Adelman-McCarthy et al.\ 2006). The SDSS data is released as DRSN1
(Sako et al.\ 2005) and DRsup (Adelman-McCarthy et al.\ 2006). The
area covered by the LAS in both $Y$- and $J$-bands in the SDSS
Southern Survey area is 106 \degt. In this area $\sim$950\,000 objects
are detected in both $Y$ and $J$ with a signal-to-noise ratio in $Y$
$\ge 7$.

We matched the UKIDSS DR1 data with the catalogues from the SDSS DR5,
using a search radius of 2\,arcsec. The colours of objects are
computed using the SDSS PSF magnitudes and the UKIDSS $\it apermag3$
magnitudes which are computed using a diameter of 2\,arcsec, corrected
for apertures losses on a pointing by pointing basis.

\section{Quasar selection}

\begin{figure}
  \includegraphics[width=8cm]{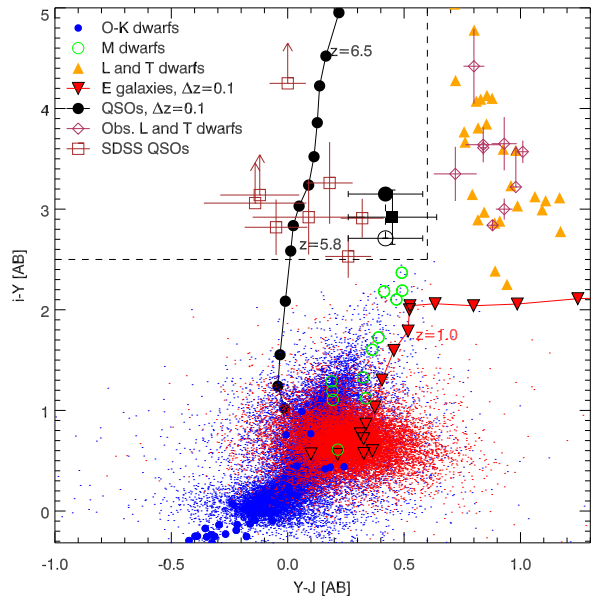}
  \caption{$i-Y$ vs $Y-J$ diagram in the AB magnitude system
    illustrating simulated colours of stars, elliptical galaxies and
    quasars (from Hewett et al.\ 2006) and observed colours from
    UKIDSS and SDSS. Also plotted are data from 20\,deg$^2$
    from the UKIDSS Early Data Release. The cloud of blue points are
    objects classified as stars, red points are objects classified as
    galaxies. The region in colour-colour space occupied by
    high-redshift quasars is outlined by the dashed lines. Quasar ULAS
    J0203+0012 is plotted as open circle with error bar (SDSS DR5
    $i$-band photometry) and as solid circle (co-added $i$-band
    photometry). The solid square is the M dwarf ULAS
    J022905.12+005506.5.}
  \label{iyj}
\end{figure}

Candidate high-redshift ($z>5.8$) quasars down to a signal-to-noise
ratio of $\sim$7 in $Y$, corresponding to a median depth of $Y_{\rm AB}=20.4$,
were selected on the basis of their blue $Y-J$ UKIDSS colour and,
either, (i) red $i-Y$ colours from our combined UKIDSS-SDSS catalogue,
or, (ii) absence in the SDSS $i$- and/or $z$-band catalogue. The
colour-colour diagram shown in Fig.\ \ref{iyj} demonstrates the
principles behind our selection criteria. The colour cuts used were
$i-Y>2.5$ and $Y-J<0.6$. The $Y-J$ cut is used to avoid the L and T
dwarfs. The $i-Y$ cut selects quasars above $z=5.8$ and discriminates
against foreground Galactic stars.

An additional requirement for candidate quasars was that they should
be undetected in the SDSS $u$-, $g$- and $r$-bands. We regarded an
object as undetected in SDSS, if the SDSS PSF-magnitude was fainter
than the 3$\sigma$ limiting magnitude of the SDSS $2048\times1489$
pixel ($13$\farcm$5\times9$\farcm$8$) field in which the object was
found. We made an empirical determination of the 3$\sigma$ magnitude
limit for each SDSS field from the SDSS catalogue. If an object was
detected in the $i$-band, we also required that $i-z>2.2$ following
the criterion used by Fan et al.\ (2003).

We performed various steps to remove false positives from our
candidate quasar sample. These steps included:

\begin{enumerate}
\item Removal of intra-quadrant cross-talk images. Cross-talk images
  are artifacts caused by bright stars (see Dye et al.\ 2006 for a
  detailed description). The cross-talk images appear offset from the
  bright star by multiples of 51.2 arcsec (or 128 pixels with scale
  0\farcs4) corresponding to a single read out channel on the same
  quadrant. More than 99\,per cent of the $z$-dropouts are cross-talk
  images. We used the 2MASS catalog to locate all stars in the
  surveyed area down to $J_{\rm AB}=15.5$. We then used the positions of these
  stars to remove all candidates that were located within a radius of
  5\,arcsec of multiples of 51.2 arcsec in the North/South or
  East/West direction as appropriate for the quadrant (see Dye et al.\
  2006) out to $\simeq$6 arcmin on the $Y$-band images from the 2MASS
  stars.

\item Independent repeat photometry on the SDSS images using the pixel
  data of the SDSS to measure the photometry of candidate quasars in
  the $g$-, $i$- and $z$-bands. This removed objects that do not
  appear in the SDSS catalogues but are present on the images, such as
  in the case where objects lie near perturbing brighter objects and
  the edges of the SDSS data. This photometric analysis also provided
  $3\sigma$ upper limits.

\item Co-adding of up to seven epochs of SDSS data from DRsup and
  DRSN1 taken in photometric conditions to detect objects that lie
  below the SDSS single-epoch catalogue limits in the optical $i$- and
  $z$-bands. The co-added images were obtained by averaging multiple
  single-epoch SDSS images centered on the location of the UKIDSS
  object, scaled to a common zeropoint using the DR5 catalogue.

\item Screening by eye to identify spurious candidates which had not
  been removed at any earlier stage (e.g.\ moving objects such as
  asteroids and corrupted objects near perturbing bright objects).
  UKIDSS observations in each waveband are not simultaneous and the
  time interval between consecutive observations in different
  wavebands is generally in the range 30 to 120 mins. Asteroids could
  be recognized as they move between the observations in the different
  UKIDSS bands. An object that appears on more than one band and
  shifts by $\ga1$\,arcsec\,hr$^{-1}$ could be identified as moving,
  down to the flux limit we considered.

\end{enumerate}

These steps reduced the number of candidates from $>$5000, the
majority of which were cross-talk artifacts, down to 10, the follow-up
observations of which are described next.

\section{Candidate follow-up observations}

\begin{table}
  \centering
  \label{mags}
  \caption{Photometry of the $z=5.86$ quasar, ULAS J020332.38+001229.2. Magnitudes are in the AB system.}
  \begin{tabular}{llcc}
    \hline
    Band     &   Source  & mag [AB] & Date \\
    \hline
    $i$ & SDSS DR5 & $>23.1$ & 2002-09-05 \\
    {} & Co-add (DRSN1, DRsup) & 23.7$\pm$0.2 & 2002--2005 \\
    $z$ & SDSS DR5 & 21.2$\pm$0.3 & 2002-09-05 \\
    {} & Co-add (DRSN1, DRsup) & 20.9$\pm$0.1 & 2002--2005\\
    $z_{NTT}$ & NTT/EMMI & 20.69$\pm$0.06 & 2006-07-18 \\
    $Y$ & UKIDSS & 20.40$\pm$0.12 & 2005-11-26 \\
    $J$ & UKIDSS & 19.98$\pm$0.10 & 2005-11-26 \\
    $K$ & UKIDSS & 19.21$\pm$0.08 & 2005-12-09 \\
    \hline
  \end{tabular}
\end{table}

\subsection{ESO NTT imaging observations}

Candidates were observed with filter $\sharp611$ in the ESO Multi-Mode
Instrument (EMMI, Dekker, Delabre \& Dodorico 1986), at the 3.6-m New
Technology Telescope (NTT)\footnote{Based on observations made with
  ESO Telescopes at the La Silla Observatory under programme ID
  077.A-0807}, for 900\,s on the night beginning 2006 August 18. The
filter/detector combination is referred to here as $z_{NTT}$ and is
similar to the SDSS $z$-band. Using the measured CCD sensitivity
curve, and the filter transmission curve, we followed the procedure
used by Hewett et al.\ (2006) to establish the colour relation
$z_{NTT}=z-0.05(i-z)$ (in AB magnitudes), for dwarf stars, O to M.
Photometry in $i$ and $z$ of SDSS DR5 sources in the field was
converted to $z_{NTT}$, and the brightness of the source was measured
via relative photometry using a fixed aperture size. For a typical
quasar at $z=5.9$ the difference between $z_{NTT}$ and $z$ is at most
0.05\,mag.

With the $z_{NTT}$-band photometry eight of the 10 observed objects
were too blue in $i-z$ and too red in $z-J$ to be considered as likely
quasars (see Table A.1 for the names of these rejected candidates).
Such objects appear to be the main contaminant in our search. Due to
photometric errors, the $z$-band magnitudes are scattered to a
brighter value, making the objects appear redder than $i-z>2.2$ and
blue in $z-J$.

The two remaining candidates, where the $z_{NTT}$
magnitude confirmed the co-added SDSS $z$-band magnitude, were
selected for spectroscopic follow-up.

\subsection{ESO VLT spectroscopic observations}

Spectra of the two candidates were obtained on 2006 September 1 and 2
using the FOcal Reducer/ low dispersion Spectrograph 2 (FORS2,
Appenzeller et al.\ 1998) on the 8.2-m Very Large Telescope (VLT)
Antu\footnote{Based on observations made with ESO Telescopes at the
  Paranal Observatory under programme ID 077.A-0310}. The candidates
were observed through the 600z holographic grism with a 1\farcs0 wide
long slit. Individual exposures were 600\,s duration. The pixels
were 2$\times$2 binned to decrease the readout time and noise, giving
a spatial scale 0\farcs25 pixel$^{-1}$ and a dispersion of
1.62\,\AA\,pixel$^{-1}$. The nights were photometric with seeing
around 0\farcs5. For the wavelength calibration, exposures of He,
HgCd and Ne arc lamps were obtained. The rms of the wavelength
calibration was better than 0.2\,\AA. The spectra cover the
wavelength range $\lambda\lambda$7370--10200. The spectrophotometric
standard star GD50 (Oke 1990) was observed with a 5\,arcsec slit for
relative flux calibration, and the absolute calibration was determined
using the observed $Y$-band flux. Total exposure times were 1200\,s
for ULAS J020332.38+001229.2 (hereafter ULAS J0203+0012) and 2400\,s
for ULAS J022905.11+005506.5.

\begin{figure}
  \includegraphics[width=8cm]{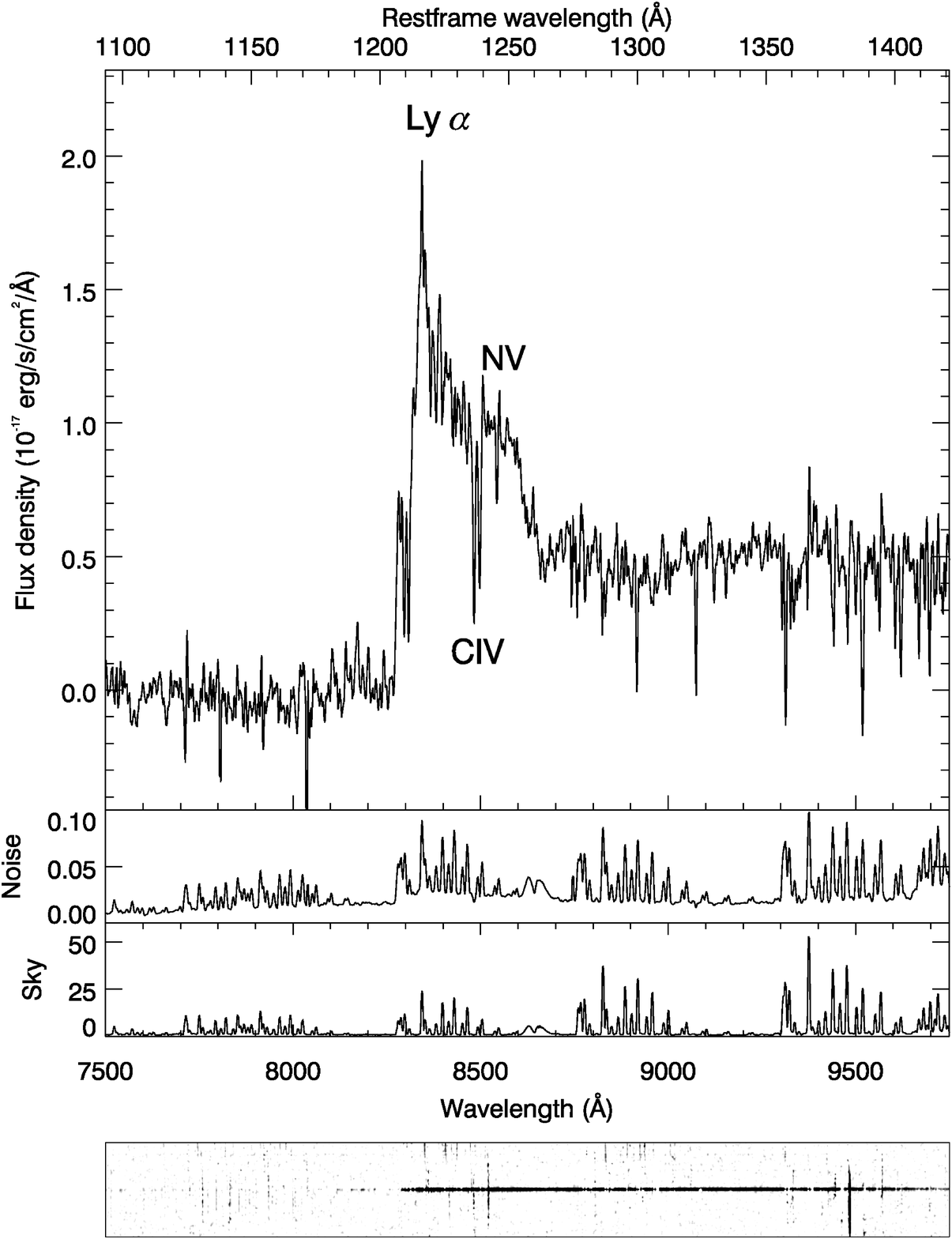}
  \caption{Discovery spectrum of the $z=5.86$ quasar ULAS
    J0203+0012. Below the object a noise and a sky spectrum are
    plotted, both in units of $10^{-17}$
    erg\,s$^{-1}$\,cm$^{-2}$\,\AA$^{-1}$. The spectra are boxcar
    averaged over three pixels. The two-dimensional spectrum is shown
    at the bottom. The Ly$\alpha$ (1216\,\AA) and N\,{\sevensize{V}}
    (1240\,\AA) emission lines are marked. The object shows the
    characteristic continuum decrement across the Ly$\alpha$ emission
    line, with very little flux blueward of the break. Around
    8500\,\AA\ a C\,{\sevensize{IV}} $\lambda\lambda$1548,1551
    absorption doublet can be seen in the spectrum. The 'absorption
    lines' at 8917 \AA, 9315 \AA\ and 9518 \AA\ are sky subtraction
    artifacts, and the absorption near 9073 \AA\ is caused by a bad
    pixel.}
  \label{spectrum}
\end{figure}

\section{Results}

\begin{figure}
  \includegraphics[width=7.25cm]{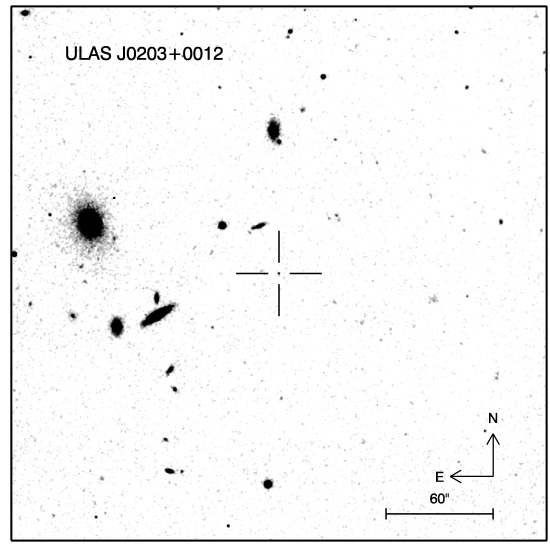}
  \caption{Finding chart for ULAS J0203+0012. The chart is taken from
    the $Y$-band image and is 5\arcmin$\times$5\arcmin\ in size. The
    image is boxcar averaged over 3$\times$3 pixels. The quasar is
    marked by a cross-hair.}
  \label{findingchart}
\end{figure}

Fig.\ \ref{spectrum} shows the spectrum of ULAS J0203+0012. The
spectrum is characterized by a broad, asymmetric emission line and a
continuum break around 8350\,\AA. We interpret these features as the
Ly$\alpha$ emission line and the continuum break resulting from
Ly$\alpha$ forest absorption at a redshift of $z\simeq5.86$. The
N\,{\sevensize{V}}~1240 emission line is blended with the Ly$\alpha$
line. The C\,{\sevensize{IV}}~1549 emission line would lie at
$\sim$10630\,\AA, outside the wavelength range of the spectrum. We
determined the redshift of the quasar by fitting a Gaussian to the red
part of the Ly$\alpha$ line. This gives a redshift of
$z=$5.864$\pm$0.003. The peak of the emission is at 8342\,\AA\ giving
a redshift of $z=5.862$. The rest-frame equivalent width of the
Ly$\alpha$\,+\,N\,{\sevensize{V}} emission line is 64\,\AA\
(uncorrected for Ly$\alpha$ forest absorption), which is comparable to
other $z\sim6$ quasars (e.g.\ Fan et al.\ 2004). A finding chart of
the quasar is shown in Fig.\ \ref{findingchart} and the magnitudes are
given in Table 1. It should be stressed that ULAS J0203+0012
is too faint in the optical passbands to be present in the SDSS DR5
catalogue. The 3$\sigma$ $z$-band magnitude listed in Table
1 is derived using aperture photometry on the SDSS pixel
image at the location of the UKIDSS source.

The second candidate, ULAS J022905.12+005506.5, turned out to be a
star, probably an M dwarf.

\section{Discussion}

\begin{figure}
  \includegraphics[width=7.5cm]{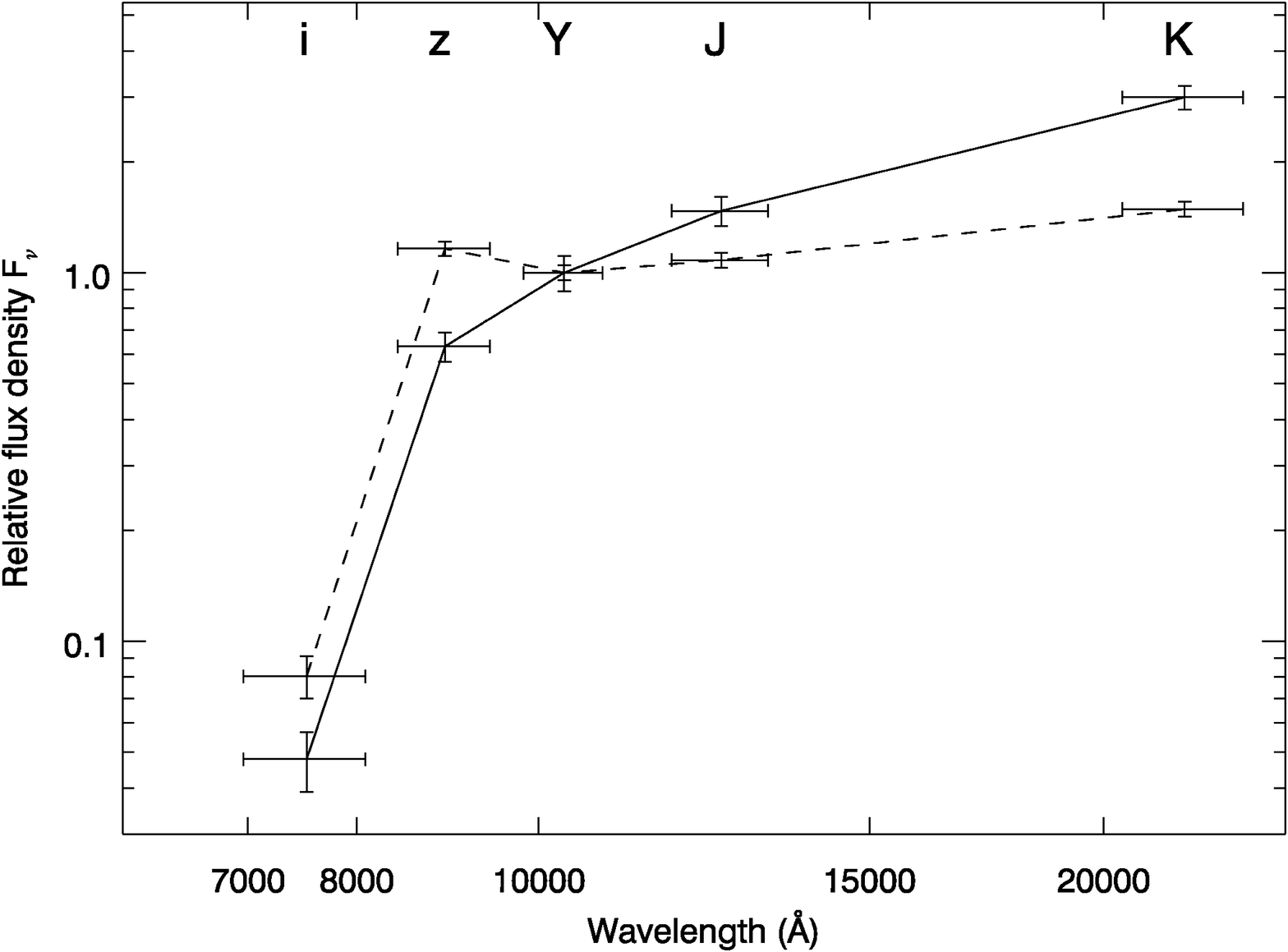}
  \caption{Spectral energy distribution of ULAS J0203+0012 (solid
    line) and a composite of three SDSS quasars with $5.8<z<6.0$.}
  \label{sed}
\end{figure}

In Fig.\ \ref{sed} we compare the optical and infrared magnitudes of
ULAS J0203+0012 with a composite of three $5.8<z<6.0$ SDSS quasars
which were observed during the science verification of WFCAM. Both
spectral energy distributions (SEDs) are scaled to $F_{\nu} = 1$ at
the centre of the $Y$-band at 10305\,\AA. ULAS J0203+0012 is
significantly redder in the infrared than the SDSS quasar
composite. If we approximate the SEDs in the infrared by a power law
with slope $\alpha$ ($F_{\nu}\propto\nu^{\alpha}$) then the slopes for
ULAS J0203+0012 are $\alpha=-1.44\pm0.17$ between the $Y$ and $K$ and
$\alpha=-1.25\pm0.21$ between $J$ and $K$. The SDSS quasars have
average slopes of $\alpha=-0.52\pm0.09$ and $\alpha=-0.56\pm0.12$
between $Y$ and $K$ and $J$ and $K$ respectively. This is consistent
with the average slope of $\alpha=-0.57$ that was found for a sample
of 45 SDSS quasars with $3.6 < z < 5.0$ by Pentericci
et al.\ (2003).

ULAS J0203+0012 is significantly fainter than the SDSS quasars. The
absolute magnitude at 1450\,\AA\ in the rest-frame is $M_{1450}=
-26.2$, nearly a magnitude fainter than that of the SDSS composite
($M_{1450}=-27.1$). However, because ULAS J0203+0012 is
redder, the extrapolated absolute $B$-band magnitude is similar to
that of the SDSS composite, $M_B=-27.9$ and $M_B=-27.7$ respectively.

\section{Conclusions}

We have reported initial results of a survey program using UKIDSS LAS
data to discover high-redshift quasars. Our results from 106 \degt\
demonstrate that the federated UKIDSS and SDSS data can be used to
carry out such a survey. The analysis and follow-up strategy we have
described is relatively efficient and should be applicable to the
entire UKIDSS LAS survey area of 4000\,deg$^2$.

Follow-up observations of the UKIDSS DR1 quasar candidates in the SDSS
Stripe 82 region are now complete. It is difficult to draw any strong
conclusions regarding the space density of quasars from one object but
we can estimate the number of quasars we would expect in the UKIDSS
survey by extrapolating the $z=6$ luminosity function of Fan et al.\
(2004) to fainter magnitudes. The density of high-redshift quasars as
found in the SDSS is $\Phi(M_{1450}<-26.7)=6\times10^{-10}$ Mpc$^{-3}$
(Fan et al.\ 2004) at a redshift of $z\sim6.1$. We assume that the
comoving density of quasars evolves with redshift as
log($\Phi$)$\propto -0.43z$ (Schmidt, Schneider \& Gunn 1995).

Number density predictions are calculated for the $Y$-band,
corresponding to a rest-frame wavelength of $\sim$1450\,\AA\ at
$z=6.1$, for which no correction for the presence of Ly$\alpha$
emission or the Ly$\alpha$ forest is required. Integrating down to
$Y_{\rm AB}=20.4$ (the median $7\sigma$ limit of the DR1 data) we expect one
quasar in $\sim$130 \degt\ in the redshift range $5.8<z<6.5$. This is
consistent with the one object, ULAS J0203+0012, that was found in 106
\degt. We can estimate in the same way the surface density of $z>6.5$
quasars in the LAS and we find one object per $\sim$390 \degt\ down to
$Y_{\rm AB}=20.4$. The UKIDSS LAS two-year plan (Dye et al.\ 2006) will cover
around $2100$ \degt. We expect this area to contain $\sim22$ new
$z>5.8$ quasars (excluding six quasars already discovered by SDSS), of
which about six would be at $z>6.5$. 

Above a redshift of $z\sim7.5$ the Ly$\alpha$ line shifts out of the
$Y$-band. Potentially quasars above this redshift could be discovered
as $Y$-band dropout objects. However, the predicted number density for
such quasars is very low ($<1$ quasar in 2100 \degt).

\section*{Acknowledgments}
We acknowledge the contributions of the staff of UKIRT, in particular
Andy Adamson, to the implementation UKIDSS survey and the Cambridge
Astronomical Survey Unit and the Wide Field Astronomy Unit in
Edinburgh for processing the UKIDSS data. We thank G.\ Miley and
F.\ Maschietto for their help in obtaining the VLT spectra, and the
referee M.\ Strauss for his valuable comments. This work is based
in part on data obtained as part of the UKIRT Infrared Deep Sky
Survey. The United Kingdom Infrared Telescope is operated by the Joint
Astronomy Centre on behalf of the U.K. Particle Physics and Astronomy
Research Council.

\appendix
\section{}

\begin{table}
  \centering
  \label{candidates}
  \caption{Quasar candidates that were rejected after the follow-up 
    observations. Using the $z_{NTT}$-band photometry (Sect.\ 4.1) these 
    eight objects are too blue in $i-z$ ($i-z<2.2$) and too red in $z-J$
    to be considered as likely quasars.}
  \begin{tabular}{l}
    \hline
    UKIDSS name \\
    \hline
    ULAS J030352.55+003334.1 \\
    ULAS J003852.24$-$005324.0 \\
    ULAS J022119.09$-$010905.9 \\
    ULAS J223845.14+011132.8 \\
    ULAS J021753.93+003906.9 \\
    ULAS J223457.11+005136.6 \\
    ULAS J011650.20+011532.2 \\
    ULAS J021047.00$-$001419.6 \\
    \hline
  \end{tabular}
\end{table}

\label{lastpage}


\begin{thebibliography}{16}
\expandafter\ifx\csname natexlab\endcsname\relax\def\natexlab#1{#1}\fi

\bibitem[{Adelman-McCarthy et al.}{2006}]{ade06} 
Adelman-McCarthy J.~K. et al., 2006, submitted to ApJS

\bibitem[{Appenzeller} {et~al.}(1998)]{app98}
{Appenzeller} I. et al. 1998, The Messenger, 94, 1

\bibitem[{Casali et al.}{2006}]{cas06}
Casali M. et al., 2006, A\&A in press

\bibitem[{Dekker, Delabre, \& Dodorico}{1986}]{1986SPIE..627..339D} 
Dekker H., Delabre B., Dodorico S., 1986, SPIE, 627, 339 

\bibitem[{Dye} {et~al.}(2006)]{dye06}
{Dye} S. et al. 2006, MNRAS, 372, 1227 

\bibitem[{Fan et al.}{2000}]{fan00} 
Fan X. et al., 2000, AJ, 120, 1167 

\bibitem[{Fan et al.}{2003}]{fan03} 
Fan X. et al., 2003, AJ, 125, 1649 

\bibitem[{Fan} {et~al.}{2004}]{fan04}
Fan X. et al., 2004, AJ, 128, 515

\bibitem[{Fan} {et~al.}{2006}]{fan06}
Fan X. et al., 2006, AJ, 132, 117

\bibitem[{{Gnedin} \& {Fan}(2006)}]{gne06}
Gnedin N.~Y., Fan X., 2006, ApJ, 648, 1

\bibitem[{Hewett et al.}{2006}]{hew06} 
Hewett P.~C., Warren S.~J., Leggett S.~K., Hodgkin S.~T., 2006, MNRAS, 
367, 454 

\bibitem[{{Kauffmann \& Haehnelt}(2000)}]{kh00}
Kauffmann G., Haehnelt M., 2000, MNRAS, 311, 576

\bibitem[{{Lawrence} {et~al.}(2006)}]{law06}
{Lawrence} A. et al., 2006, submitted to MNRAS (astro-ph/0604426)

\bibitem[{{Oke}(1990)}]{oke90}
{Oke} J.~B., 1990, AJ, 99, 1621

\bibitem[{Page} {et~al.}(2006)]{pag06}
{Page} L. et al., 2006, submitted to ApJ (astro-ph/0603450)

\bibitem[{Pentericci et al.}{2003}]{pen03} 
Pentericci L. et al., 2003, A\&A, 410, 75 

\bibitem[{{Sako} {et~al.}(2005)}]{sak05}
{Sako} M. et al. 2005, astro-ph/0504455

\bibitem[{{Schmidt} {et~al.}(1995){Schmidt}, {Schneider}, \& {Gunn}}]{sch95}
{Schmidt} M., {Schneider} D., {Gunn} J., 1995, AJ, 110, 68

\bibitem[{Skrutskie} {et~al.}(2006)]{skr06}
{Skrutskie} M.~F. et al., 2006, AJ, 131, 1163

\bibitem[{Storrie-Lombardi et al.}{2001}]{sto01} 
Storrie-Lombardi L.~J., Irwin M.~J., McMahon R.~G., Hook I.~M., 2001, 
MNRAS, 322, 933 

\bibitem[{Warren} {et~al.}(2006)]{war06}
{Warren} S.~J. et al., 2006, MNRAS in press (astro-ph/0610191)

\bibitem[{York} {et~al.}(2000)]{yor00}
{York} D.~G. et al., 2000, AJ, 120, 1579

\end{thebibliography}
\end{document}